\documentclass[aip,reprint]{revtex4-2}
\usepackage{color}
\usepackage{graphicx}
\usepackage{dcolumn}
\usepackage{bm}
\usepackage{amsmath}
\usepackage[caption=false]{subfig} 
\usepackage{braket}
\usepackage[english]{babel}

\draft 

\begin{document}

\title{Clocked Quantum-dot Cellular Automata Circuits Tolerate Unwanted External Electric Fields}
\thanks{This work was funded by the Office of Naval Research under grant N00014- 20-1-2420.}

\author{Peizhong Cong}
	\email{joe\_cong1@baylor.edu}
\author{Enrique P.  Blair}%
	\email{Enrique\_Blair@baylor.edu.}
\affiliation{Department of Electrical and Computer Engineering, Baylor University, Waco TX}%

\date{\today}

\begin{abstract}
Quantum-dot cellular automata (QCA) may provide low-power, general-purpose computing in the post-CMOS era. A molecular implementation of QCA features nanometer-scale devices and may support  \(\sim\)THz switching speeds at room-temperature. Here, we explore the ability of molecular QCA circuits to tolerate unwanted applied electric fields, which may come from a variety of sources. One likely source of strong unwanted electric fields may be electrodes recently proposed for the write-in of classical bits to molecular QCA input circuits. Previous models have shown that the input circuits are sensitive to the applied field, and a coupled QCA wire can successfully transfer the input bit to downstream circuits despite strong applied fields. However, the ability of other QCA circuits to tolerate an applied field has not yet been demonstrated. Here we study the robustness of various QCA circuits by calculating their ground state responses in the presence an applied field. To do this, a circuit is built from several QCA molecules, each described as a two-state system. A circuit Hamiltonian is formed and diagonalized. All pairwise interactions between cells are considered, along with all correlations. An examination of the ground state shows that these QCA circuits may indeed tolerate strong unwanted electric fields. We also show that circuit immunity to the dominant unwanted field component may be obtained by choosing the orientation of constituent molecules. This suggests that relatively large electrodes used for bit write-in to molecular QCA need not disrupt the operation of nearby QCA circuits. The circuits may tolerate significant electric fields from other sources, as well.
\end{abstract}

\maketitle 

\section{Introduction}
\label{sec:orgd19acb8}

Quantum-dot Cellular Automata (QCA) is a promising post-CMOS, general-purpose computing paradigm may support energy-efficient, high-speed computing.\cite{LentTougawPorodBernstein:1993} The elementary QCA device is called a cell, a system of mobile charges on a set of coupled quantum dots. A cell's electronic configuration encodes a classical bit, and the inter-dot quantum tunneling of charge enables device switching. Coulomb coupling between neighboring QCA cells enables arrangements of cells on a surface to function as logic circuits that can perform general classical computation. Mixed-valence molecules may provide nanometer-scale cells with redox centers as quantum dots.\cite{LentScience2000,molecularQCAelectronTransfer,mQCA_00}  Molecular QCA may support ultra-high switching speeds and room-temperature operation.\cite{molecularQCAelectronTransfer} These characteristics make molecular QCA a desirable candidate for beyond-CMOS, general purpose computing.


In this paper, we explore the extent to which molecular QCA computational circuits tolerate unwanted local electric fields, \(\vec{E}\). While such fields could come from a variety of sources, one source may apply significantly strong electric fields: nearby electrodes designed to write classical bits to molecular QCA circuitry.\cite{QCAeFieldWriteIn-IEEE,cong_robust_2021} It has been shown that electrodes much larger than the QCA molecules can be used for bit write-in to input circuits. These electrodes also may immerse nearby circuits in strong, unwanted fringing electric fields. It has been demonstrated that the binary wires and shift registers coupled to electric-field input circuits tolerate significant unwanted applied fields.\cite{cong_robust_2021} In this study, we apply similar methodology to basic QCA interconnective circuits and logic circuits to examine their ability to function under unwanted electric fields.

It will be shown here that molecular QCA circuits can tolerate relatively strong unwanted applied fields. In section \ref{background}, we give the general background of molecular QCA. In section \ref{model}, we describe a two-state device model along with a treatment of an \(M\)-cell circuit in its full, \(2^M\)-dimensional Hilbert space. Then, in section \ref{results}, results are presented indicating the robustness of the basic QCA interconnects and logical building blocks under applied interfering electric fields. We conclude with discussion about minimizing the potential for unwanted fields to disrupt the operation of QCA logic and interconnects.

\section{Background \label{background}}
\label{sec:org697e993}

\begin{figure}[htbp]
\centering
\includegraphics[width=\linewidth]{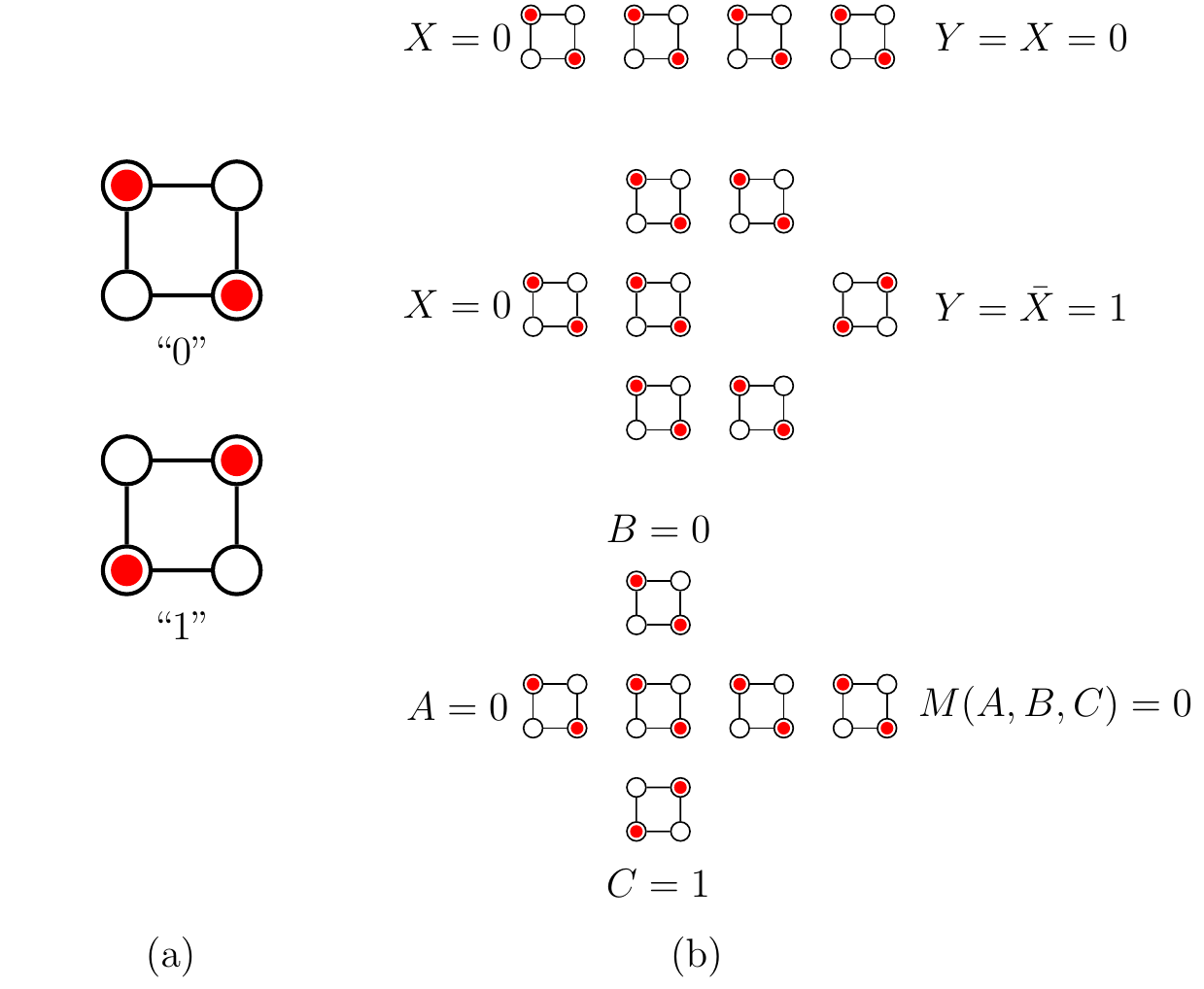}
\caption{(a) QCA four-dot cell in both states and (b) basic logic built with QCA four-dot cells. \label{fig:fourdotcell}}
\end{figure}

Fig.\ \ref{fig:fourdotcell}(a) schematically shows the binary device states of a four-dot QCA cell with two mobile electrons: ``0'' and ``1.'' The black circles represent quantum dots, red discs represent mobile electrons, and black lines indicate tunneling paths for the mobile charge. QCA cells interact through inter-cell Coulomb coupling. Fig.\ \ref{fig:fourdotcell}(b) shows basic logic that can be implemented with the four-dot QCA cells. In a binary wire (top), an input bit \(X\) is copied successively from one cell to the next through nearest-neighbor interactions, so that the output bit is \(Y=X\). An inverter (middle) uses diagonal, next-nearest-neighbor interactions to invert input bit \(X\) to obtain the output \(Y=\bar{X}\). A  majority gate (bottom) outputs the bit \(M(A, B, C)\) in the majority of the three inputs, \(A\), \(B\), and \(C\).

Various implementations of QCA have been demonstrated experimentally. QCA were first implemented using metal islands patterned on an insulating oxide.\cite{Snider1998,LeadlessQCA} Later, GaAs/AlGaAs heterostructure semiconductor dots\cite{Smith2003} and silicon dots\cite{mitic_demonstration_2006} were used to build QCA cells. Dangling bonds on a H-passivated silicon surface provide dots for atomic-scale QCA robust at room temperature.\cite{WolkowQCA_Silicon} Basic QCA circuits such as binary wires\cite{Orlov1999} and majority logic have been demonstrated.\cite{Snider1999} Clocked devices, such as latches\cite{clocked_qca_latch} and shift registers\cite{kummamuru_power_2001} also have been demonstrated. Clocking is important because it enables quasi-adiabatic device operation, provides for the control of the direction and timing of calculations, and provides power gain to restore weakened bits.\cite{toth_quasiadiabatic_1999,Timler2002}

\begin{figure}[htbp]
\centering
\includegraphics[width=\linewidth]{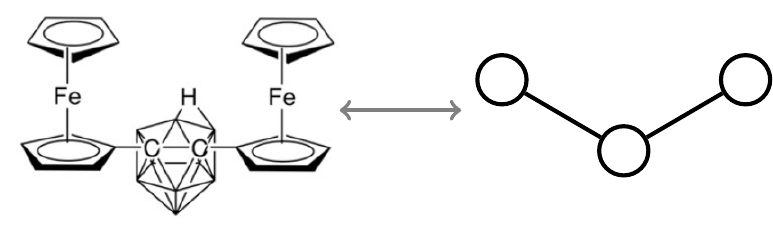}
\caption{A zwitter-ionic nidocarborane molecule provides a three-dot QCA cell. \label{fig:molecule}}
\end{figure}

This paper focuses on a molecular implementation of QCA, in which redox centers on mixed-valence (MV) molecules function as quantum dots.\cite{LentScience2000,mQCA_00,2003-JACS-mQCA} This implementation may provide nanometer-scale devices with densities approaching \(10^{14} \; \text{cm}^{-2}\) and device switching speeds hundreds or even thousands of times faster than modern transistors.\cite{molecularQCAelectronTransfer} Cationic molecules have been synthesized and tested as QCA devices. Field-driven device switching has been observed in molecular QCA.\cite{mQCA_STM_studies} Even single-molecule switching has been observed.\cite{AFM_switch_2020} Charge-neutral, MV zwitterionic candidates are a promising class of molecule, since their synthesis avoids the random placement of counterions near the QCA devices.\cite{2013_zwitterionicQCA_DQD} This is desirable, since stray charge affects device switching characteristics.\cite{2005-Counterions-2} The first zwitterionic MV QCA candidate to be synthesized and tested was the zwitterionic nido-carborane molecule of Figure Fig.\ \ref{fig:molecule}.
\cite{ZwitterionicNidoCarborane} Two ferrocene groups  and the carborane cage each provide a quantum dot. The central cage also traps the fixed neutralizing charge, which provides a controllably-localized counter-ion to balance the mobile charge.

\begin{figure}[htbp]
\centering
\includegraphics[width=\linewidth]{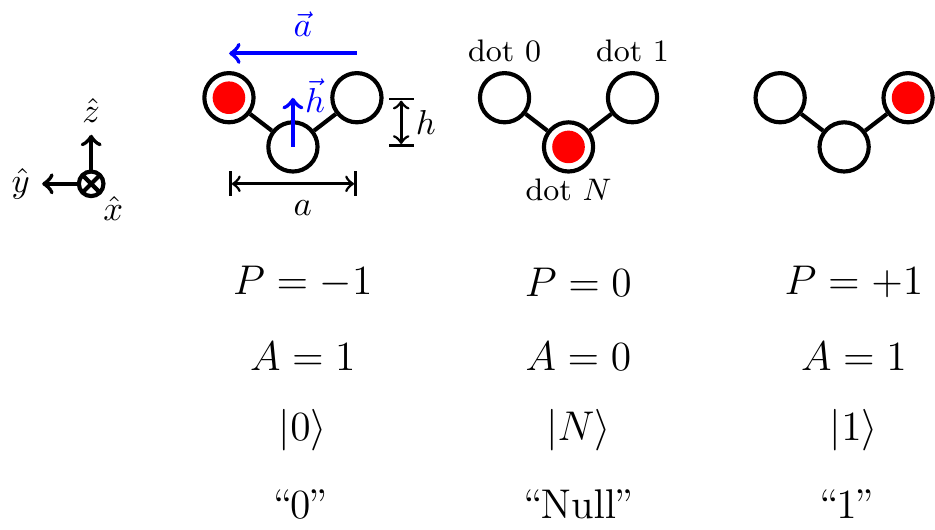}
\caption{A single mobile electron on a system of three coupled quantum dots provides a clocked three-state QCA cell. \label{fig:threedotcell}}
\end{figure}

In this paper, circuits modeled are formed using three-dot molecules similar to the zwitterionic nido-carborane.  The molecules used here are assumed to have a mobile electron and a neutralizing hole trapped on the central dot. Figure \ref{fig:threedotcell} schematically depicts the three localized electronic states that define the device states: ``0,'' ``1,'' and ``Null.'' The mobile electron is represented as a red disc, and the trapped neutralizing hole is not shown. The dots are labeled 0, 1, and \(N\), with dots 0 and 1 designated ``active'' dots, since they are used to encode a bit. The active dots are separated by distance \(a = \left| \vec{a} \right|\). With dot \(N\) on the device plane (\(z=0\)), the active dots are elevated above dot \(N\) in the plane \(z= h\).  The vectors \(\vec{a}\) (pointing from dot 0 to dot 1) and \(\vec{h} =h \hat{z}\) will determine how an applied field biases the device states.

The lines connecting the dots in the schematic indicate a tunneling path. Here, transitions between the active states (``0'' and ``1'') occur via an intermediate the ``Null'' state, and direct transitions between active states  are suppressed. This will enable electric-field clocking of QCA cells using the \(z\)-component of an applied electric field.\cite{Hennessy2001,BlairLentArchitecture,2018_clock_topologies} A clocking field \(E_z \hat{z}\) with sufficiently strong \(E_z < 0\) can over come the affinity of the mobile electron for the neutralizing charge on dot \(N\), driving the cell from the ``Null'' state to the active state biased by neighbor interactions. As long as an adequate activating field is applied, the bit is latched. Reversing the sign of \(E_z\) draws the electron into dot \(N\) and into the ``Null'' state regardless of neighbor interactions.

Recent work also has shown that one component of the applied electric field along \(\vec{a}\)\textemdash say, \(E_y \hat{y}\), for example\textemdash may be used to program bits onto input circuits. This is feasible even if interconnecting binary wires or clocked shift registers Coulomb-coupled to the input elements are immersed in the same input field.\cite{QCAeFieldWriteIn-IEEE,cong_robust_2021} This is made possible by the sensitivity of an input circuit segment to the input \(E_y \hat{y}\) and the insensitivity of the wire/shift register to \(E_y \hat{y}\).

The use of large electrodes in the vicinity of molecular devices raises questions about the ability of QCA logic to function under unwanted applied fields, \(\vec{E} (\vec{r})\). Can a local \(\vec{E} (\vec{r})\) disrupt logic circuits, and under what conditions can the logical devices return correct computational results? These are the primary concerns of this paper.

\begin{figure}[htbp]
\centering
\includegraphics[width=\linewidth]{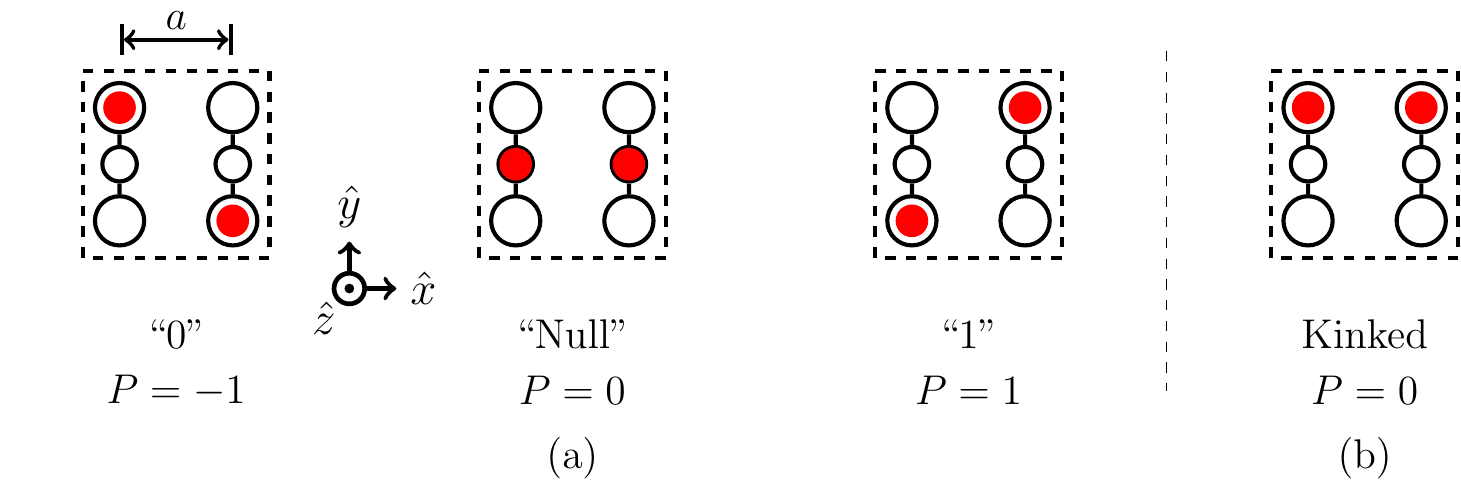}
\caption{QCA three-dot cell pairs function as a six-dot cell. \label{fig:cellgroup}}
\end{figure}

In this study of QCA circuits, we focus on pairs of three-dot molecules, as in Figure \ref{fig:cellgroup}. Here, two three-dot cells separated by a distance \(a\) form a pair. Each individual molecule is half of a six-dot cell, which has the same binary states as the four-dot cell along with an additional ``Null'' state, as in Figure \ref{fig:cellgroup}(a). Here, we define the polarization, \(P\) of a cell pair in terms of \(P_L\) and \(P_R\), the polarizations of the left and right three-dot half-cells, respectively: \(P_{\text{pair}} = (P_1 - P_0)/2\).  While unpaired three-dot molecules interact strongly with \(\vec{E} (\vec{r})\) as dipoles, cell pairs interact with \(\vec{E}\) more weakly as quadrupoles. This is precisely the reason binary wires and shift registers tolerate immersion in an unwanted \(\vec{E}\), yet the state of an input segment is highly sensitive to \(\vec{E}\). Nonetheless, a strong \(\vec{E}\) may drive a kink within a cell pair, with \(P=0\) as in Figure \ref{fig:cellgroup}. The weak quadrupole interaction of cell pairs with \(\vec{E}\) enables QCA logic to tolerate applied unwanted fields.

A useful energy scale is the kink energy, \(E_k\), the cost of a kink in a cell pair relative to the energy of states ``0'' or ``1.'' This may be calculated from electrostatics as \(E_k = q_e^2(1-1/\sqrt{2})/4\pi \varepsilon_0 a\), where \(\varepsilon_0\) is the vacuum permittivity, and \(q_e\) is the fundamental charge. Then, \(E_o\), the field strength required to inject a kink in an isolated cell pair, becomes a useful scale for the strength of the applied field:
\begin{equation}
E_o = \frac{E_k}{q_e a} = \frac{q_e}{4\pi \varepsilon a^2} \left( 1 - \frac{1}{\sqrt{2}} \right) \; .
\end{equation}
For three-dot cells with \(a=1\; \text{nm}\), \(E_o = 0.417 \; \text{V/nm}\).

The separation between cell pairs in a circuit may be chosen to optimize desirable device interactions and to minimize an undesirable device interaction known as population congestion.\cite{PopCongest2020} Here, Coulomb repulsion between mobile charges in the plane of the active dots \(z = h\) increases the strength of the clock \(E_z\) required to active cells.

\section{Model \label{model}}
\label{sec:org096c45a}

The model for a QCA circuit in the presence of an applied electric field is based on a two-state reduction of a three-state device description of each cell. This reduces the dimension of the circuit's Hilbert space, allowing the modeling of larger circuits with the two-state devices. 

\subsection{Isolated QCA three-dot cell}
\label{sec:org4c4a0ec}

The Hamiltonian of a single, clocked QCA three-dot cell can be written as
\begin{align}
\label{eq:hamiltonian:3dc}
\hat{H}=&-\gamma\left(\hat{P}_{0, N} + \hat{P}_{1, N} + \hat{P}_{N,
1} + \hat{P}_{N, 0}\right) \nonumber \\
&+ \frac{\Delta}{2}\hat{Z} + \left(V_c + V_N - E_a\right)\hat{P}_N.
\end{align}
Here, we make use of transition operators, \(\hat{P}_{\alpha,\beta}\equiv\ket{\alpha}\bra{\beta}\),  which describe \(\ket{\beta} \rightarrow \ket{\alpha}\) transitions. Also, projection operators \(\hat{P}_\alpha\equiv\ket{\alpha}\bra{\alpha}\) are used. The first term describes tunneling between the active states and the null state with energy \(\gamma\). In the second term, a detuning, \(\Delta\), characterizes the bias between the two active states: \(\Delta=\braket{1|\hat{H}|1} - \braket{0|\hat{H}|0}\), and \(\hat{Z} \equiv \hat{P}_1 - \hat{P}_0\).  The detuning \(\Delta = \Delta_{\text{neigh}} + \Delta_{\vec{E}} + \Delta_o\) includes contributions from (1) neighbor electrostatic interactions, \(\Delta_{\text{neigh}}\); (2) the local applied field, \(\Delta_{\vec{E}} = -q_e \vec{E} \cdot \vec{a}\), and (3) any chemical bias \(\Delta_o\) between the active device states.\cite{2021-Asymmetric-QCA} Here, it will be assumed that \(\Delta_o = 0\).  The device polarization \(P\) is the expectation value of \(\hat{Z}\). The third term describes the bias of the ``Null'' state relative to the energy of the active states. \(V_N\) describes interactions with neighboring cells, and \(V_c = -q_e \vec{E} \cdot \vec{h} \) is a function of both the applied clocking field and molecule geometry.  \(E_a\) is the affinity of the mobile electron for the null dot and the fixed neutralizing charge it hosts. \(E_a\) is a molecular property unique to each particular QCA candidate.

\subsection{Reduced two-state QCA cell}
\label{sec:org07cb893}

The above three-state description of a QCA molecule was previously used to model clocked molecular QCA input circuits with electric field inputs, including all pairwise intercellular interactions and all intercellular correlations.\cite{cong_robust_2021} Since the quantum state space for a circuit model grows exponentially with the number of constituent molecules, \(M\), this treatment becomes prohibitively expensive for anything but the smallest of circuits. Since an \(M\)-cell circuit has a \(3^M\)-dimension state space with a \(3^{M}\times3^{M}\) Hamiltonian, only circuits with \(M\leq8\) were found to be feasible on modern high-performance computing systems.

 To explore the behaviors of logic circuits, it is necessary to use \(M>8\). Therefore, we reduce the three-state description to a two-state model for a QCA device. This may be achieved by assuming a strong clocking field, \(E_z \hat{z}\), that suppresses the population of mobile charge on the Null dot. Thus, a three-state vector \(\ket{\psi} = c_0 \ket{0} + c_N \ket{N} + c_1 \ket{1} \) has \(c_N = \braket{N|\psi} \rightarrow 0\) so that \(\ket{\psi} \simeq c_0 \ket{0} + c_1 \ket{1}\) . This reduction of each QCA cell's state space reduces the dimension the circuit Hilbert space to \(2^{M}\) and makes it practicable to model circuits with up to \(M=14\) molecules.

We use the following two-state Hamiltonian for a QCA cell with negligible ``Null'' state population:
\begin{align}
\label{eq:hamiltonian:r3dc}
\hat{H}=&-\gamma_\text{eff}\hat{\sigma}_x + \frac{\Delta}{2}\hat{\sigma}_z,
\end{align}
where \(\hat{\sigma}_x = \hat{P}_{0,1} + \hat{P}_{1,0}\) and \(\hat{\sigma}_z = \hat{P}_1 - \hat{P}_0\) are Pauli operators. \(\hat{\sigma}_z\) is analogous to the three-state operator \(\hat{Z}\), so that a cell's polarization is obtained by taking the expectation value of \(\hat{\sigma}_z\): \(P = \braket{\hat{\sigma}_z}\). The first term of Equation (\ref{eq:hamiltonian:r3dc})  describes tunneling between the active states of the molecule, with effective tunneling energy \(\gamma_{\text{eff}}\).

\(\gamma_{\text{eff}}\) may be obtained via second-order perturbation theory: \(\gamma_{\text{eff}}\) is half of the energy of the avoided crossing separating the ground state from the first excited state. To estimate this value, we consider an isolated QCA cell and the three-state \(\hat{H}\) from Equation \ref{eq:hamiltonian:3dc}, with no bias between active states (\(\Delta = V_N = 0\)). It may be shown that
\begin{equation}
\gamma_{\text{eff}} = \frac{\sqrt{\left(V_c - E_a\right)^{2}+8\gamma^{2}} - \left(V_c - E_a\right)}{4} \; .
\label{eq:gamma:eff}
\end{equation}
A reasonable value of \(\gamma = 100~\text{meV}\) may be assumed, along with distances \(a=1\; \text{nm}\) and \(h = 0.5 \; \text{nm}\). A strongly applied clock of \(E_z \sim -10 E_o\) and \(E_a = 1\; \text{eV}\) leads to \(\gamma_{\text{eff}} = 10\; \text{meV}\) and suppresses an isolated cell's null dot population in the ground state to no more than \(\braket{\hat{P}_N} = 0.02\), indicating that the two-state reduction is reasonable. Throughout this paper, we assume \(\gamma_{\text{eff}} = 10\; \text{meV}\).

\subsection{QCA circuit Hamiltonian}
\label{sec:orgb50e3b3}

A basis \(\{\ket{\mathbf{x}}\}\) may be formed for an \(M\)-cell QCA circuit  by taking direct products of localized single-cell states:
\begin{align}
\Ket{\mathbf{x}}=\Ket{x_M  \cdots x_2 x_1}=\Ket{x_M}\cdots\Ket{x_2}\Ket{x_1},
\label{eq:basis:circuit}
\end{align}
where \(\mathbf{x}\) is the \(M\)-bit word \(x_M  \cdots x_2
x_1\), and \(\Ket{x_k} \in \{\Ket{0}, \Ket{1}\}\) is the
\(k\)-th single cell state.

The circuit Hamiltonian can be written as
\begin{align}
\label{eq:hamiltonian:circuit}
\hat{H}=\hat{H}_\text{int} + \sum_{k=1}^M \hat{H}_k,
\end{align}
where \(\hat{H}_\text{int}\) is the interaction Hamiltonian that describes the Coulomb interaction between the QCA cells; \(\hat{H}_{k}\) is the \(k\)-th single cell Hamiltonian. The interaction \(\hat{H}_{\text{int}}\) is diagonal in basis \(\{\ket{\mathbf{x}}\}\) and may be written as
\begin{align}
\hat{H}_{\text{int}} = \sum_{\mathbf{x}} U_{\mathbf{x}} \hat{P}_{\mathbf{x}}.
\label{eq:hamiltonian_interaction}
\end{align}
Here, \(U_{\mathbf{x}}\) is the electrostatic energy of interaction between all pairs of cells in the circuit, given state \(\mathbf{x}\):
\begin{align}
U_{\mathbf{x}} & = \Braket{ \mathbf{x}| \hat{H}_{\text{int}} | \mathbf{x} } \; . 
\label{eq:es_energy}
\end{align}
The interaction energies are calculated as in previous work.\cite{QCAeFieldWriteIn-IEEE,cong_robust_2021} The QCA circuit ground state is found by diagonalizing the \(M\)-cell Hamiltonian.

The research method here is to determine the conditions under which the ground state of a circuit immersed in an unwanted field \(\vec{E} (\vec{r})\) encodes the desired logical response.

\subsection{Extreme limit}
While unwanted non-clocking electric fields \(E_x \hat{x} + E_y \hat{y}\) could come from numerous external sources, an important motivation for this study is the exploration of circuit operation under fringing fields from input electrodes. The applied \(E_x \hat{x} + E_y \hat{y}\) may be related only indirectly to the input bits applied to the molecular circuits, since arbitrary transformations may have taken place between the input circuits and a given circuit. How well do QCA circuits tolerate fringing fields from input electrodes or other sources? 

\begin{figure}[htbp]
\centering
\includegraphics[width=\linewidth]{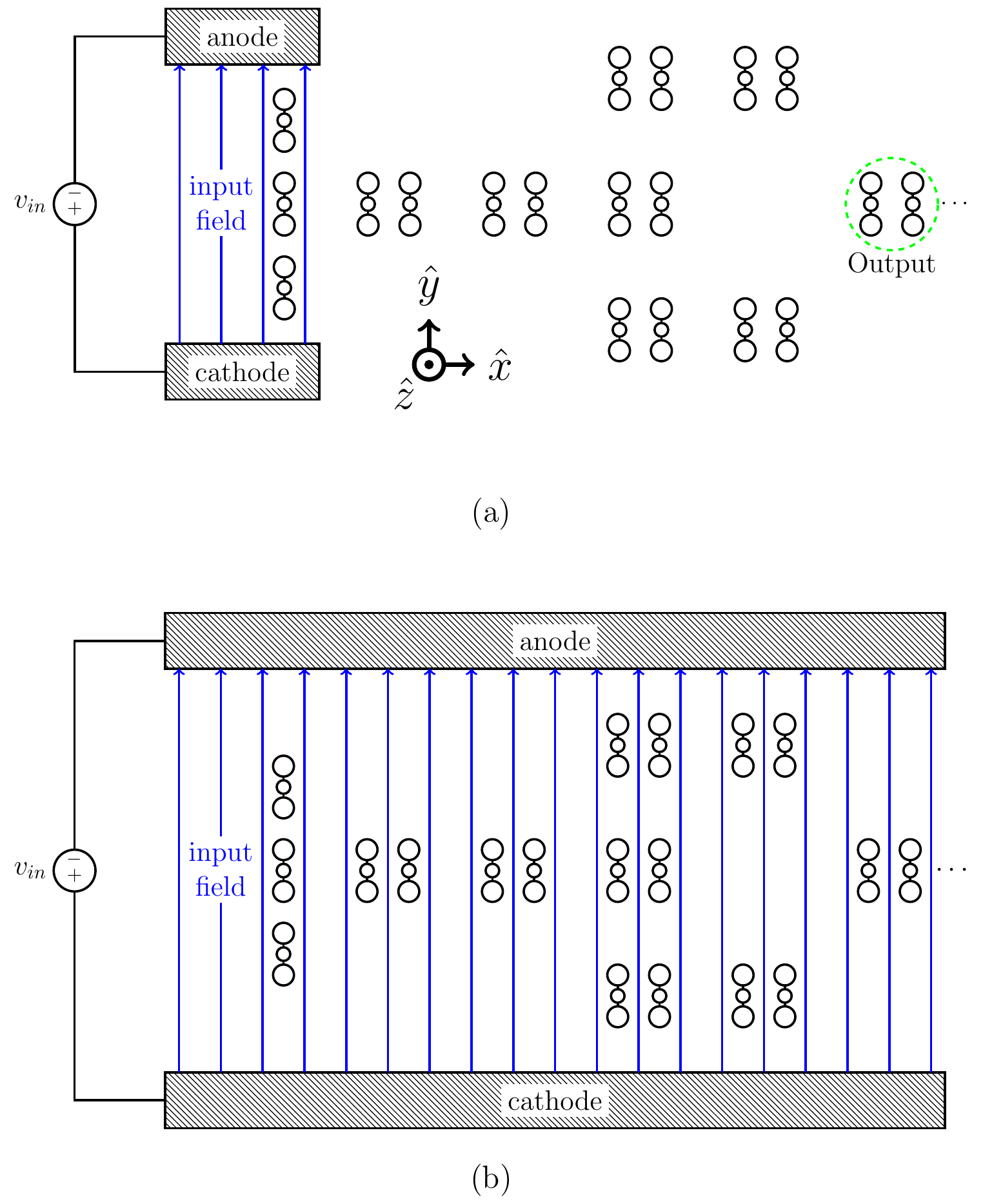}
\caption{A QCA input and inverter are shown in two limits: (a) in the ideal limit, an input field is constrained only to the three left-most cells, which are used as in input circuit; and, (b) a worst-case limit, in which the same field is applied uniformly over the input circuit and computational circuits. More realistically, a weaker fringing field from the electrodes may exist in the vicinity of the computational circuits.\label{fig:limits}}
\end{figure}

Ideally, the electrodes apply an input field, \(E_y \hat{y}\), that is constrained to the input circuits and is negligible over regions where QCA logic exists, as in Figure \ref{fig:limits}(a). In this case, the input fields do not to interfere with the molecular logic. This is feasible to the extent that nanoscale electrodes with single-molecule specificity may be fabricated.

Another extreme limit is where the input electrodes are so large that the input circuits and the QCA logic all are immersed in the same, uniform input field, as in Figure \ref{fig:limits}(b). This large-electrode limit may be viewed as a worst-case limit, since an intermediate regime is more likely and realistic: some fringing, non-clocking \(E_x \left( \vec{r}\right) \hat{x} + E_y \left( \vec{r}\right) \hat{y}\) applied over the logic will be weaker than the field applied over the input circuits. Furthermore, we assume that the \(y\)-component of the field, \(E_y \hat{y}\) is the dominant non-clocking component (\(\left| E_y\right| \gg \left|E_x \right|\)), since it is used for inputs. Finally, only input fields are considered under the assumption that these dominate other sources of unwanted fields.

In this paper, uniform fields \(\vec{E}\) are used, as in the worst-case, large-electrode limit. If the ground state of the circuits studied here can encode the desired logical output even in the large-electrode limit, it is reasonable to conclude that they will function properly in more realistic scenarios.


\subsection{Rotation of cell groups}
Molecules may be oriented to minimize or even eliminate sensitivity to the dominant non-clocking field component. For example, molecules with \(\vec{a} \cdot \hat{y} = 0\) lose sensitivity to the input field, \(E_y \left( \vec{r} \right) \).\cite{cong_robust_2021}  It will be seen that circuits tolerate strong applied \(E_y \hat{y}\) for cells, even with \(\vec{a} \cdot \hat{y} \neq 0\). Additionally, rotating constituent cell groups can add immunity to the dominant field component.

Technologies that control the position and orientation of nanometer-scale molecules with the precision envisioned here have not yet been demonstrated experimentally. Some combination of self-assembly and lithography could one day provide suitable molecular circuit layout methods.\cite{hu_high-resolution_2005,DNARaftsExperiment,2014_EBL_DNA} Layout technologies require further development, and we hope that this paper may provide some motivation for experimental efforts in this area.

\section{Results \label{results}}
\label{sec:org37fe0b8}

The results given here are for interconnective and logic circuits, since input circuits have been previously considered.\cite{QCAeFieldWriteIn-IEEE,cong_robust_2021} Driver cell pairs simulate input bits for the circuits modeled here. These bits may be shifted to the logic via interconnecting wires from upstream inputs or logical operations. Interconnective circuits are useful in performing inversions and bit erasures with a copy. Specifically, we model fan-in and fan-out circuits. Inverters and majority gates are studied, since together, these fundamental gates provide a logically complete set. 


\subsection{Fan-in}
\label{sec:org6caec54}

\begin{figure}[htbp]
\centering
    \subfloat[Fan-in circuit. \label{fig:fan-in-reg}]{
      \includegraphics[width=0.45\columnwidth]{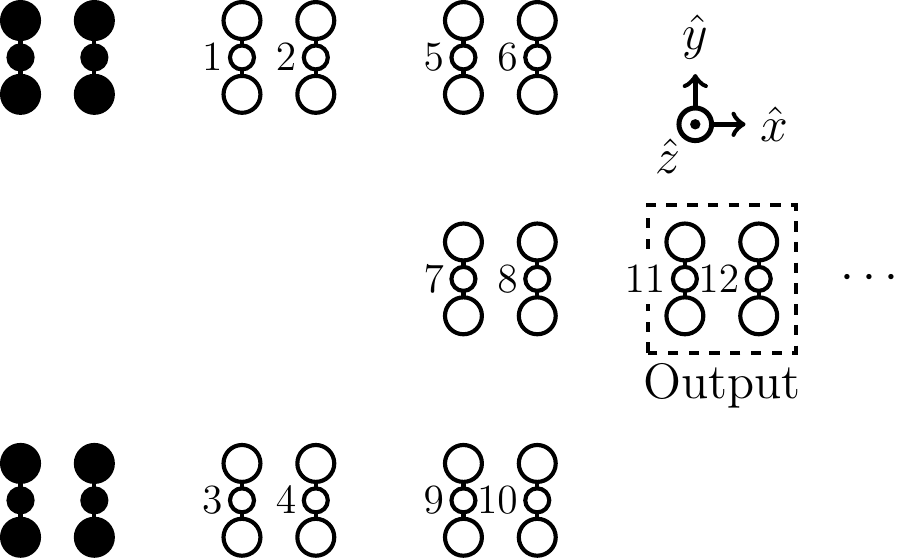}
      }
      \quad
    \subfloat[Input-insensitive fan-in.\label{fig:fan-in-imp}]{
      \includegraphics[width=0.45\columnwidth]{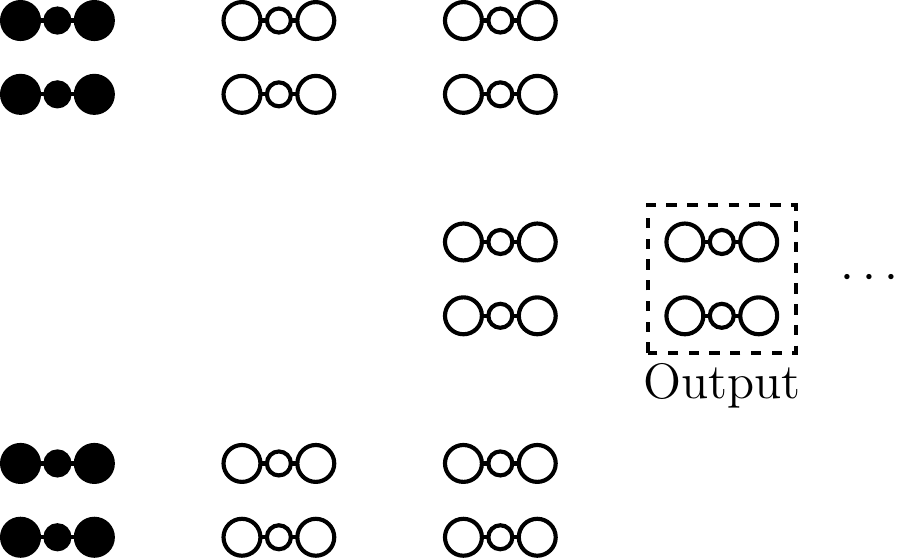}
      }
\caption{(a) A fan-in circuit is built from 12 QCA molecules, forming six cell pairs. Two driver pairs (black dots) provide two copies of the same input bit. The fan-in outputs only one copy of the same bit. (b) A 90-degree rotation of cell pairs makes the fan-in circuit insensitive to the input field component, \(E_y \hat{h}\). \label{fig:model:fanin}}
\end{figure}

A fan-in circuit, shown schematically in Fig.\ \ref{fig:fan-in-reg}, reduces two bits into one. This is useful for performing an erasure with a copy, allowing the bit energy of an erased copy\textemdash which may be significant\textemdash to be returned to the clock rather than dissipated to the environment.\cite{blair_signal_2011}

The fan-in of Fig.\ \ref{fig:model:fanin} is built from six paired QCA molecules. Two additional cell pairs (left, unlabeled and colored black) function as drivers, providing two identical bits as an input for the circuit.  Only one copy of the input bit proceeds rightward from the circuit output. The grayed out ``Null'' dots on the QCA cells indicate electron population is suppressed on these dots using a strong clock, which allows for the use of the reduced two-state model.

\begin{figure}[htbp]
\centering
\includegraphics[width=\linewidth]{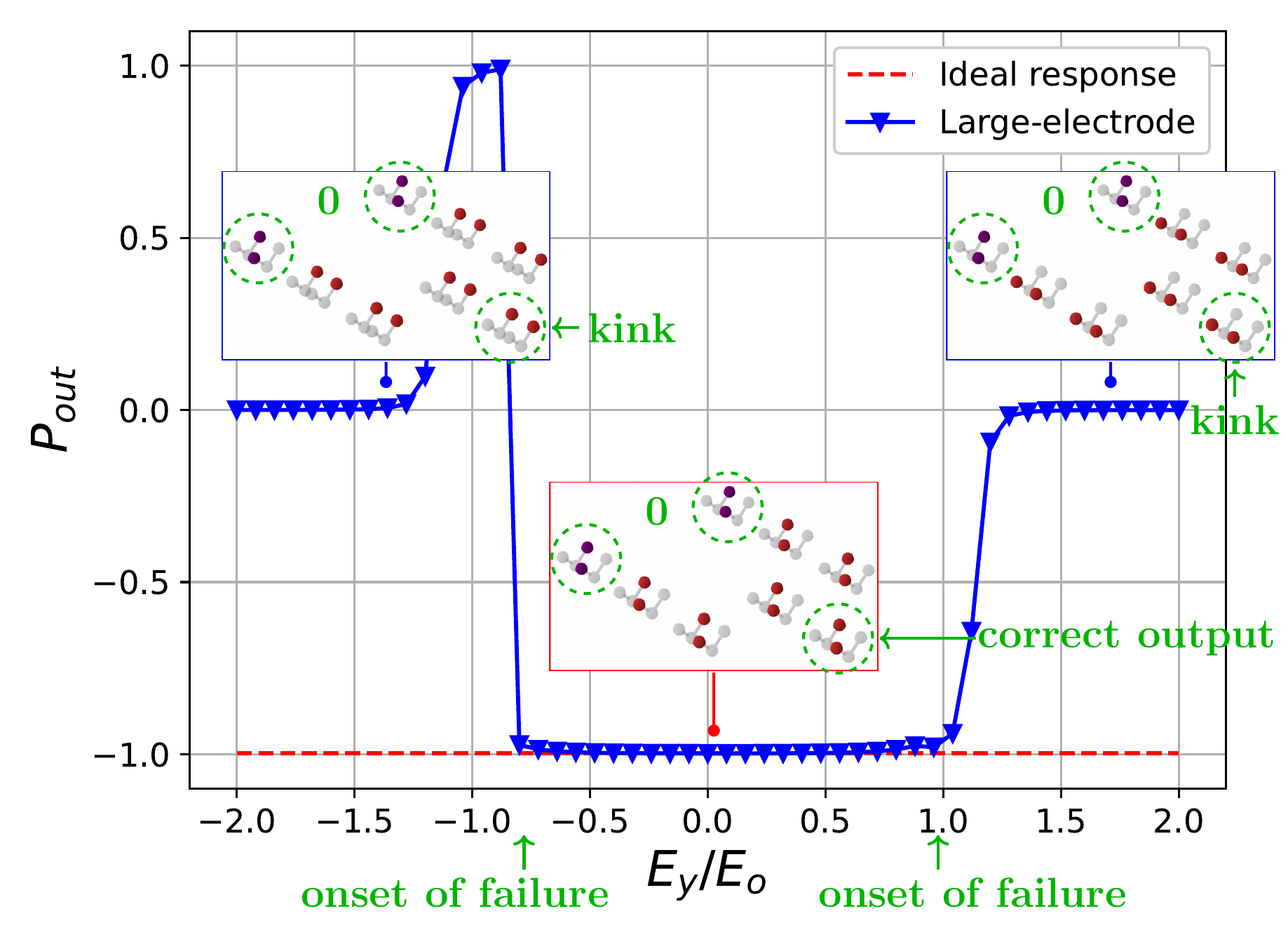}
\caption{A clocked QCA fan-in circuit responds to an input bit \(0\) in the presence of an applied input field, \(E_y \hat{y}\). \label{fig:fanin:result}}
\end{figure}

The fan-in circuit outputs the desired ground state response to an input bit 0 and a uniformly-applied input field, \(E_y \hat{y}\), as shown in Fig.\ \ref{fig:fanin:result}. 
The circuit output remains close to the ideal value (under zero non-clocking field, \(E_x \hat{x} + E_y \hat{y} = 0\)) until \(E_y \hat{y}\) is strong enough to disrupt circuit operation. The onset of failure is estimated to be \(E_y \simeq 0.8 E_o\). Stronger \(E_y \hat{y}\) introduces kinks within individual cell groups.
Since input fields \(|E_y| \ll 0.5E_o\) may be used to write input bits,\cite{cong_robust_2021} it is straight-forward write bits using input electrodes and avoid by a significant margin those field strengths which may disrupt nearby QCA logic.

Circuit failure under very strong \(E_y \hat{y}\) is facilitated by the alignment of dipoles between cells 5-10. This alignment lowers the cost of a kink in cell groups (5,6), (7,8), and (9,10). This response is asymmetric in \(E_y \hat{y}\) because the input bits favor a kink in one direction over the other. By the symmetry of the circuit, we can reason that its response to a ``1'' bit may be obtained by reflecting the data in Fig.\ \ref{fig:fanin:result} about the \(x\) axis and then about the \(y\) axis.

\begin{figure}[htbp]
\centering
\includegraphics[width=\linewidth]{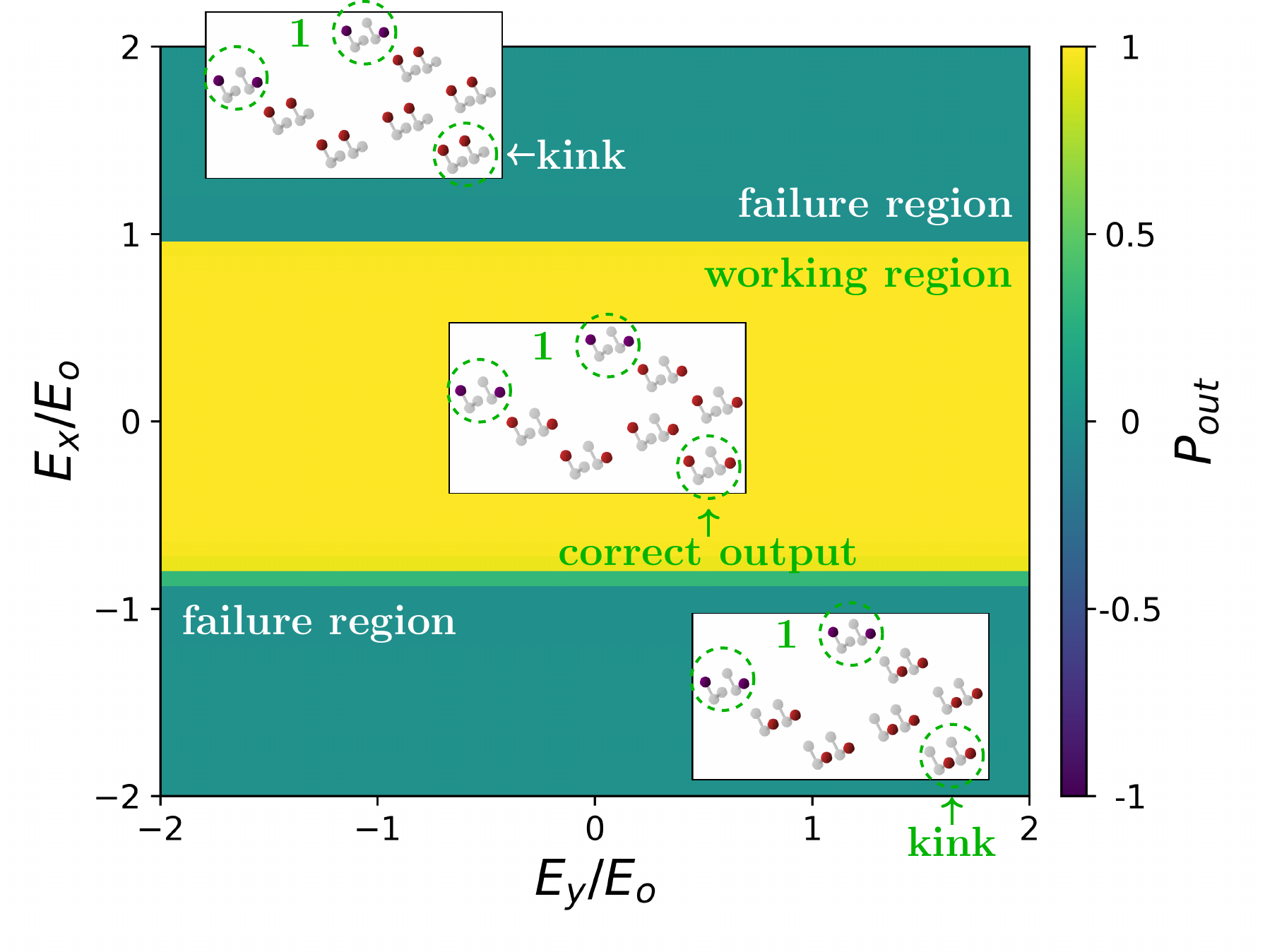}
\caption{A clocked QCA fan-in circuit with rotated cell pairs responds to bit \(1\) inputs and applied, non-clocking fields \(E_x \hat{x} + E_y \hat{y}\). \label{fig:fanin:result:rot}}
\end{figure}

Figure \ref{fig:fan-in-imp} shows a 90-degree rotation of the cell pairs forming the molecular circuit. This makes it insensitive to \(E_y \hat{y}\), since the cells have \(\vec{a} \perp \hat{y}\). However, sensitivity to \(E_x \hat{x}\) must now be considered. Fig.\ \ref{fig:fanin:result:rot} comfirms that the ground state response of the fan-in circuit with rotated cell pairs is not affected by \(E_y \hat{y}\) whatsoever: the circuit can provide the correct output (yellow region) regardless of arbitrarily-strong \(E_y \hat{y}\). Strong \(E_x \hat{x}\) may now disrupt circuit operation (green region, \(P_{\text{out}} \rightarrow 0\)), however. The circuit fails earlier for \(E_x < 0\) than for \(E_x > 0\): the circuit operates over \(0.8 E_o < E_x < 0.9 E_o\). This is because it is easier to inject kinks by pushing the mobile charge away from that of the drivers (\(E_x < 0\)) than it is to pull the mobile charge toward the like charge of the drivers (\(E_x > 0\)). 
Given the need to handle both input cases, the fan-in with rotated cells can operate in the presence of strong fields, but we should constrain the non-clocking field so that \(\left| E_x \right| \lesssim 0.8 E_o\).
This simple rotation of cell groups further improves the robustness of QCA fan-in circuits.

\subsection{Fan-out}
\label{sec:orge87535c}

\begin{figure}[htbp]
\centering
    \subfloat[Fan-out circuit. \label{fig:fan-out-reg}]{
      \includegraphics[width=0.45\columnwidth]{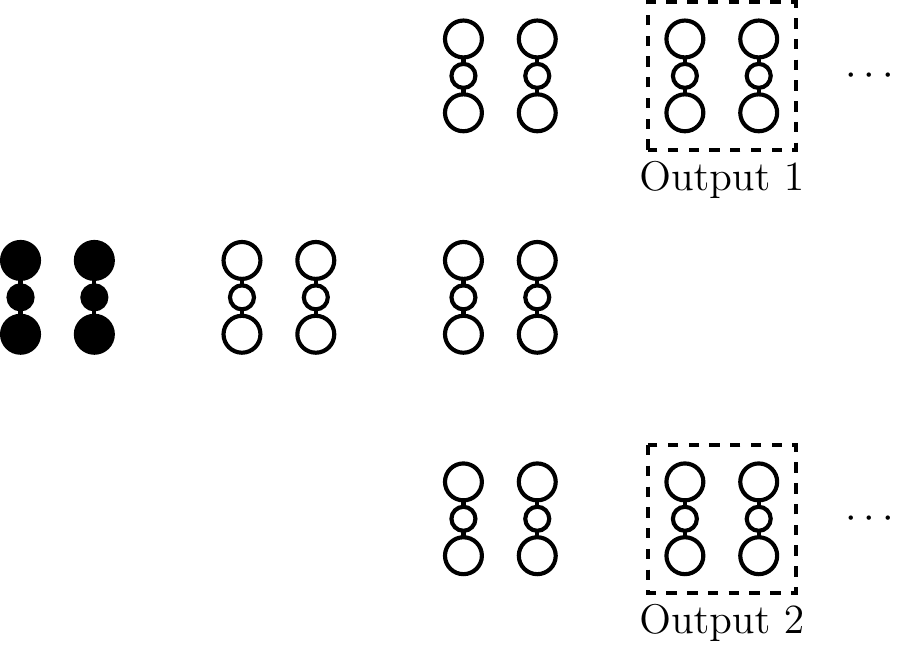}
      }
      \quad
    \subfloat[Input-insensitive fan-out. \label{fig:fan-out-imp}]{
      \includegraphics[width=0.45\columnwidth]{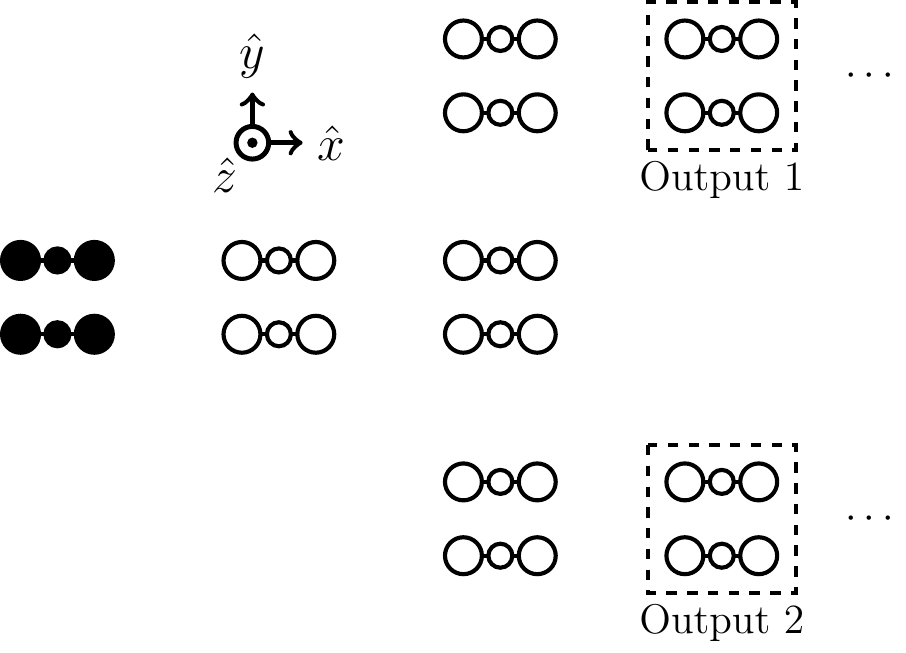}
      }
\caption{A QCA fan-out circuit is built from twelve three-dot molecules (six cell pairs). (a) One input pair (black molecules) provides an input bit, and the circuit outputs two copies. (b) The fanout's cell pairs are rotated 90 degrees for immunity to the input field \(E_y \hat{y}\). \label{fig:model:fanout}}
\end{figure}

A fan-out circuit, shown in Fig.\ \ref{fig:model:fanout}, takes a single-bit input and outputs two copies of the same bit. The circuit modeled here is built from six cell groups.


\begin{figure}[htbp]
\centering
\includegraphics[width=\linewidth]{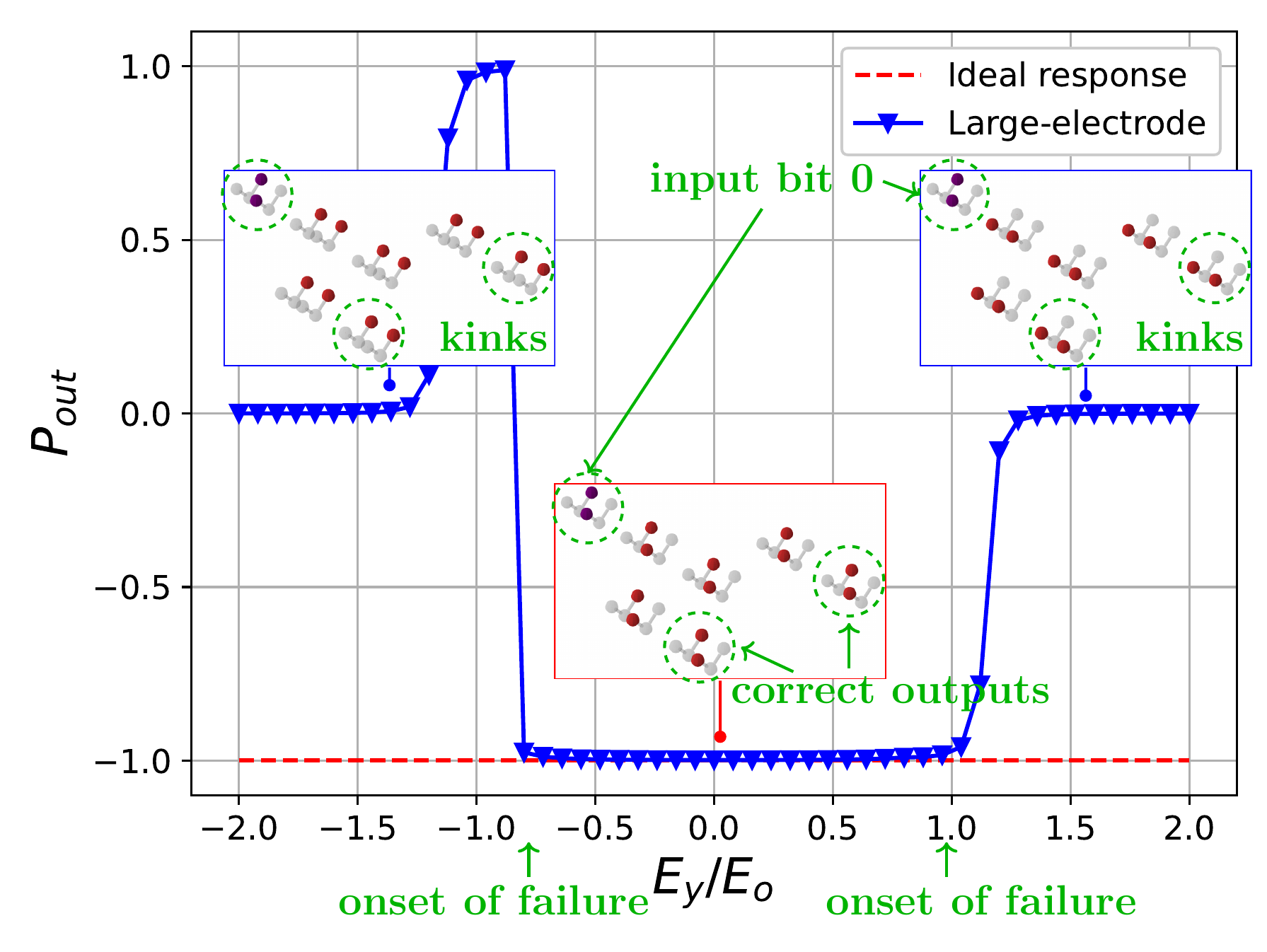}
\caption{A strongly-clocked QCA fan-out circuit responds to input bit \(0\) and an applied input field, \(E_y \hat{y}\). \label{fig:fanout:result}}
\end{figure}

The fan-out circuit of Fig.\ \ref{fig:fan-out-reg} tolerates strong, unwanted applied field component \(E_y \hat{y}\), as shown in  Fig.\ \ref{fig:fanout:result}. Here, an input bit 0 is provided,  
and the output signal plotted here is the average of the polarizations from the two output cell groups. The fan-out tolerates unwanted applied fields in the input direction up to about \(|E_y| \sim 0.8 E_o\) before the circuit response departs from the ideal response. As in the case of the fan-out, the alignment of dipoles between cells facilitates device failure, and the input bit breaks the symmetry of the circuit's response to \(E_y \hat{y}\), causing the circuit to fail in one direction of \(E_y \hat{y}\) before the other. By the symmetry of this problem, the circuit response to input bit 1 may be obtained by reflecting this result about both the \(x\) and \(y\) axes. 

\begin{figure}[htbp]
\centering
\includegraphics[width=\linewidth]{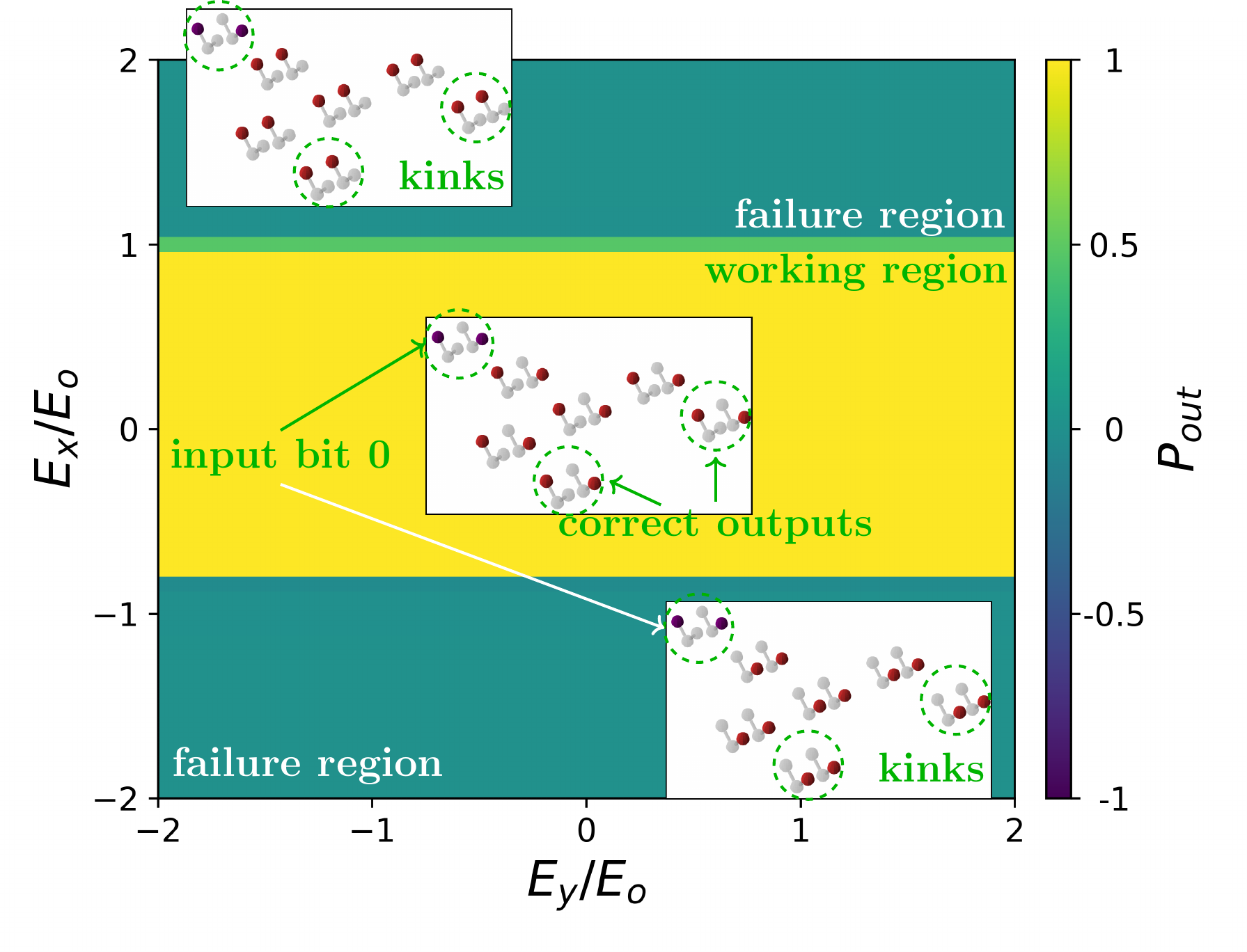}
\caption{A strongly-clocked fan-out circuit with rotated cell groups respond to bit \(1\) input and applied fileds \(E_{y}\) in the dominant direction and \(E_{x}\) in the signal direction. \label{fig:fanout:result:rot}}
\end{figure}

As in the case of fan-in, a fan-out consisting of rotated cell groups enjoys immunity to applied input fields \(E_y \hat{y}\). Fig.\ \ref{fig:fanout:result:rot} shows the ground state response of such a fan-out circuit, with both \(E_x \hat{x}\) and \(E_y \hat{y}\) components. Again, the output polarization is the average of the polarizations of the two output cell groups. The circuit correctly produces two copies of bit 0 (yellow region) regardless of \(E_y \hat{y}\); however, strong \(E_x \hat{x}\) can drive circuit failure (green region).  The region for correct circuit operation is estimated to be \(\left| E_x \right| \lesssim 0.8E_o\).

\subsection{Inverter}
\label{sec:org14d2ecc}

\begin{figure}[htbp]
\centering
    \subfloat[Inverter circuit.\label{fig:inverter-reg}]{
      \includegraphics[width=0.45\columnwidth]{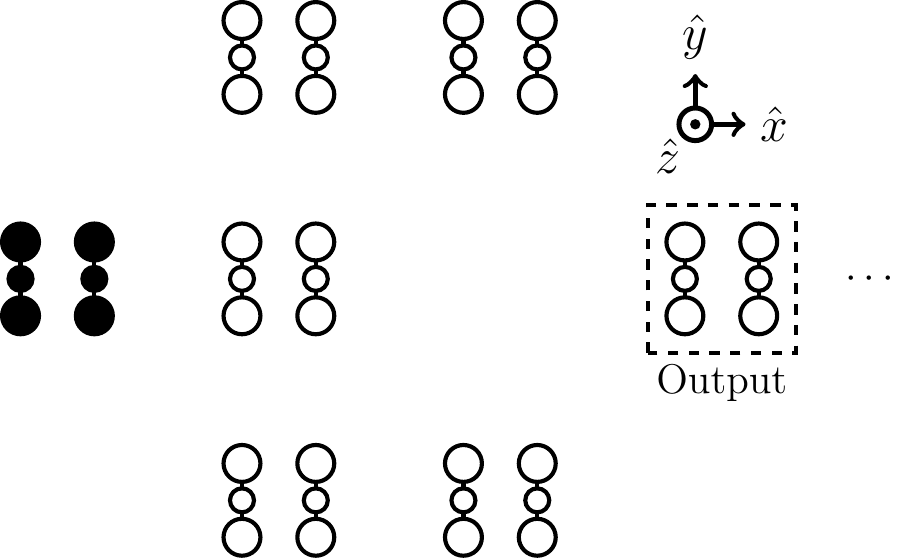}
      }
      \quad
    \subfloat[Input-insensitive inverter.\label{fig:inverter-imp}]{
      \includegraphics[width=0.45\columnwidth]{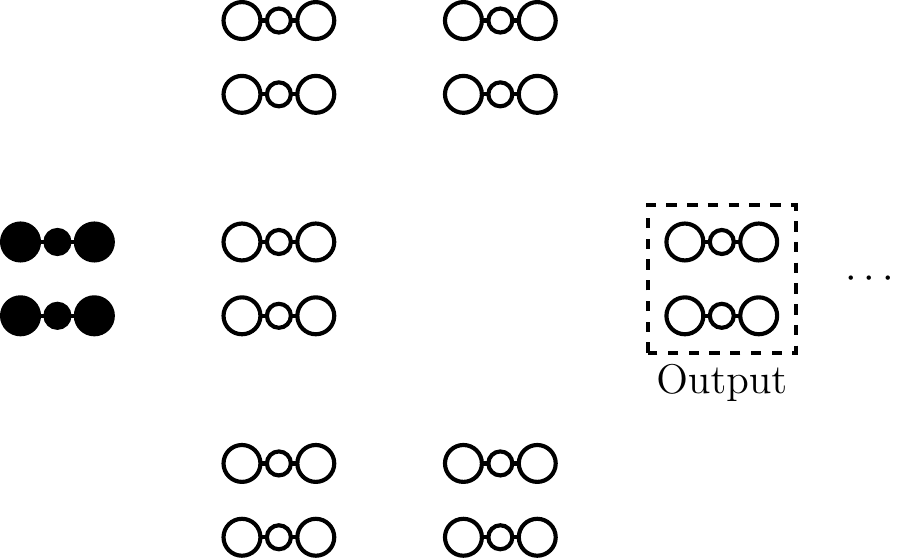}
      }
\caption{A QCA inverter is built from six cell pairs (twelve molecules). One driver pair provides an input bit, which is fanned out into two copies. These copies interact with the output pair via an inverting next-nearest-neighbor interaction. \label{fig:model:inverter}}
\end{figure}

Fig.\ \ref{fig:inverter-reg} schematically shows the simplified version of the inverter circuit that was previously shown in Fig.\ \ref{fig:limits}. In the QCA inverter, a single input bit is fanned out into two copies and inverted using a diagonal, next-nearest-neighbor coupling between the input bit copies and the output cell. While inversion could be achieved with only one input bit and next-nearest-neighbor coupling to the output, the fan-out enhances the robustness of the circuit: the second bit copy increases the Coulomb interaction with the output cell group, raising the cost of a bit error. The full fan-out is not modeled here, since it already was treated previously. The focus here is on field-driven disruption of the inverting, next-nearest neighbor coupling between the two input bit copies and the output cell.

\begin{figure}[htbp]
\centering
\includegraphics[width=\linewidth]{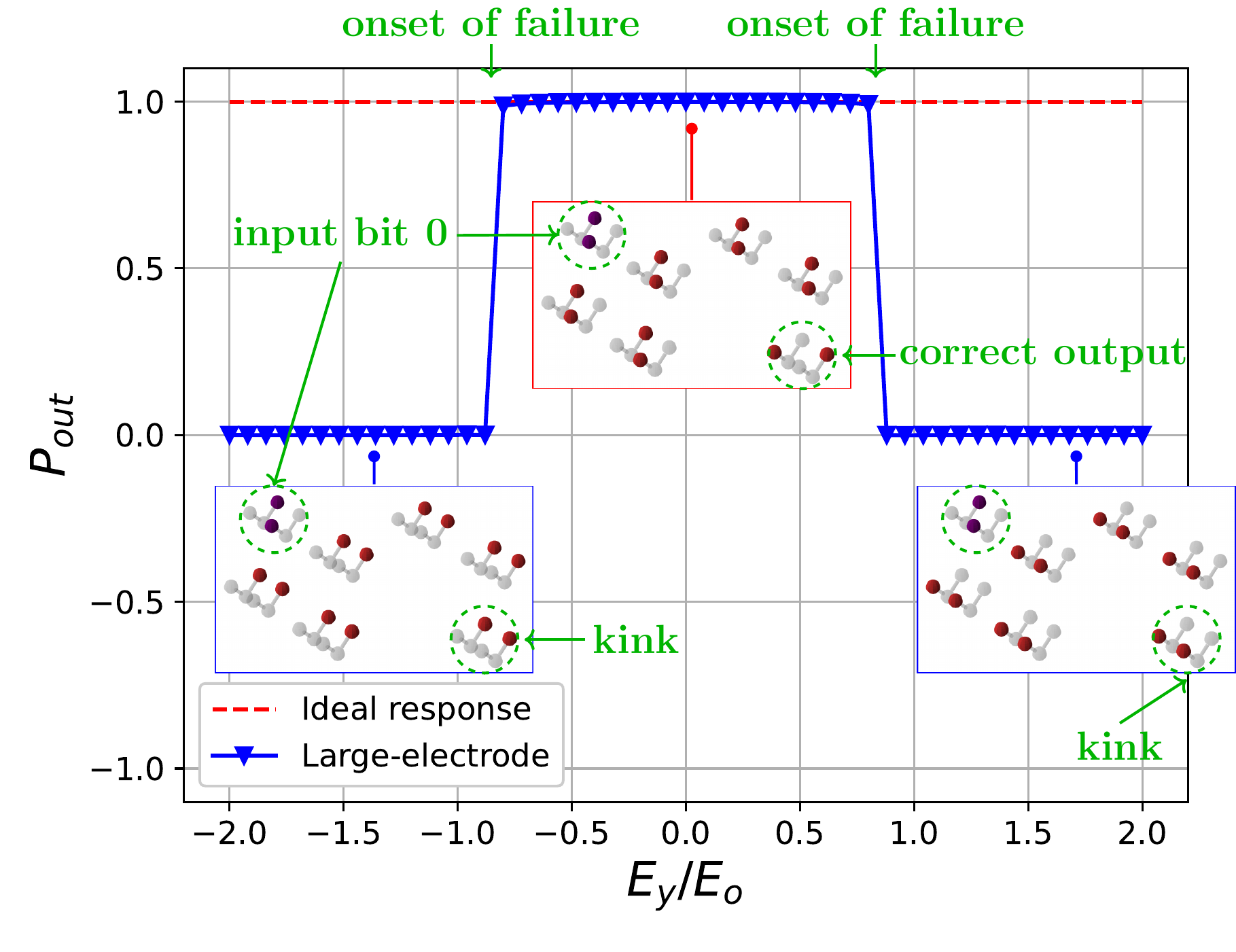}
\caption{A clocked QCA inverter responds to input bit 0 and an applied input field, \(E_y \hat{y}\).\label{fig:inverter:result}}
\end{figure}

In the presence of an unwanted field \(E_y \hat{y}\), the onset of failure for the inverter of Fig.\  \ref{fig:model:inverter} is approximately \(|E_y| \sim 0.8E_{o}\), as with other circuits. The inverter's response to \(E_y \hat{y}\) and an input bit 0 is shown in Fig.\ \ref{fig:inverter:result}. The circuit's response to bit 1 under an applied \(E_y \hat{y}\) may be obtained by reflecting the data across the \(x\) axis.

\begin{figure}[htbp]
\centering
\includegraphics[width=\linewidth]{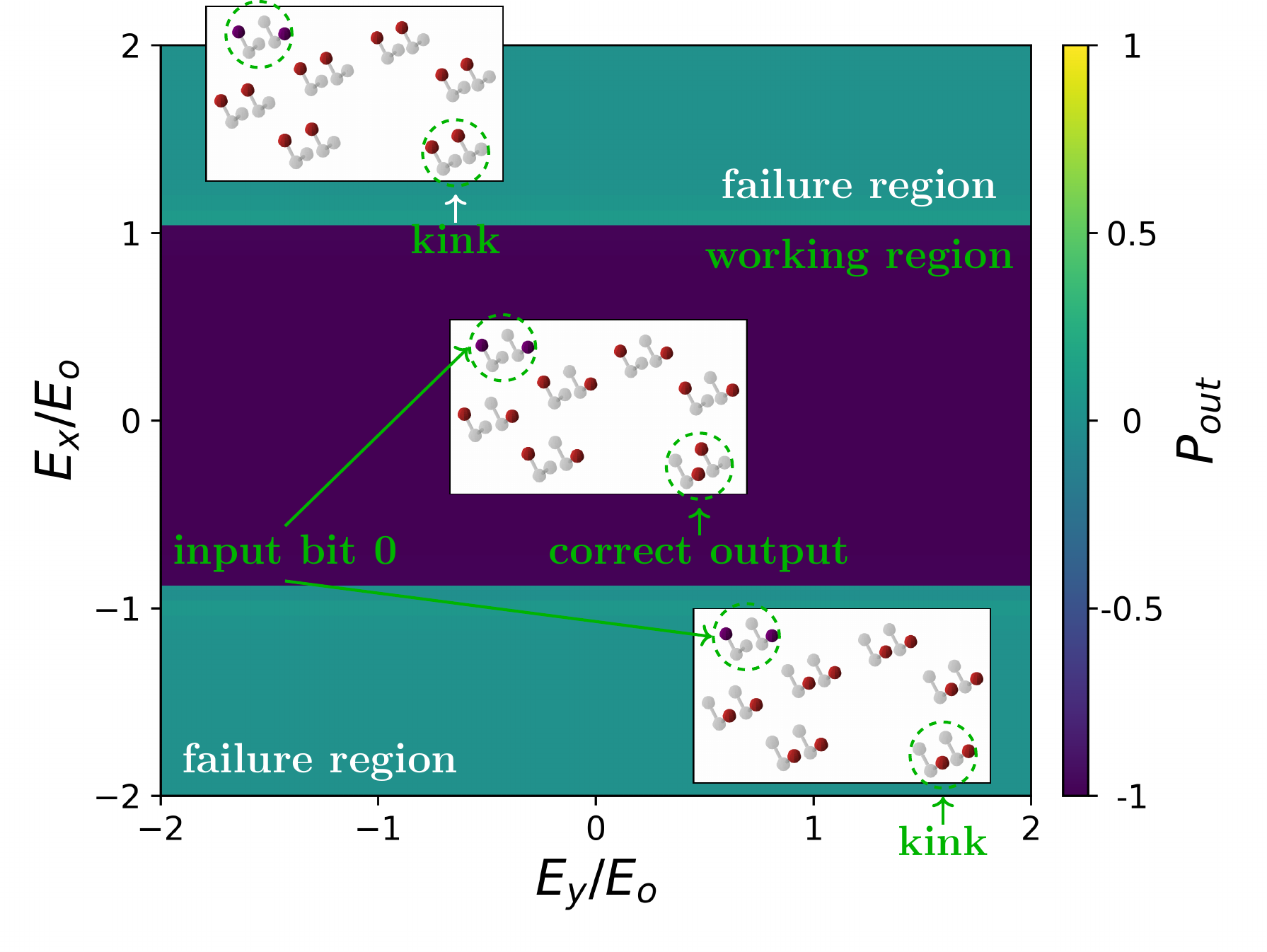}
\caption{A clocked QCA inverter with rotated cell pairs respond to input bit 1 and applied non-clocking fields \(E_x \hat{x} + E_y \hat{y}\). The circuit returns the correct bit regardless of \(E_y\) and tolerates strong \(E_x\). \label{fig:inverter:result:rot}}
\end{figure}

As with other circuits, a 90-degree rotation of the cell groups (see Figure \ref{fig:inverter-imp}) eliminates any sensitivity to the input field component, \(E_y \hat{y}\), but introduces sensitivity to \(E_x \hat{x}\). As before, the rotated cell pairs still tolerate strong field components \(E_x \hat{x}\). Fig.\ \ref{fig:inverter:result:rot} shows the ground state response of a QCA inverter with rotated cell pairs to an input bit 1 and applied non-clocking fields, \(E_x \hat{x} + E_y \hat{y}\). The blue region is where the circuit successfully inverts input bit 1 to output bit 0 (\(P_{\text{out}} = -1\)). 

\subsection{Majority Gate}
\label{sec:org9b88551}

\begin{figure}[htbp]
\centering
    \subfloat[Majority circuit. \label{fig:majority-reg}]{
      \includegraphics[width=0.45\columnwidth]{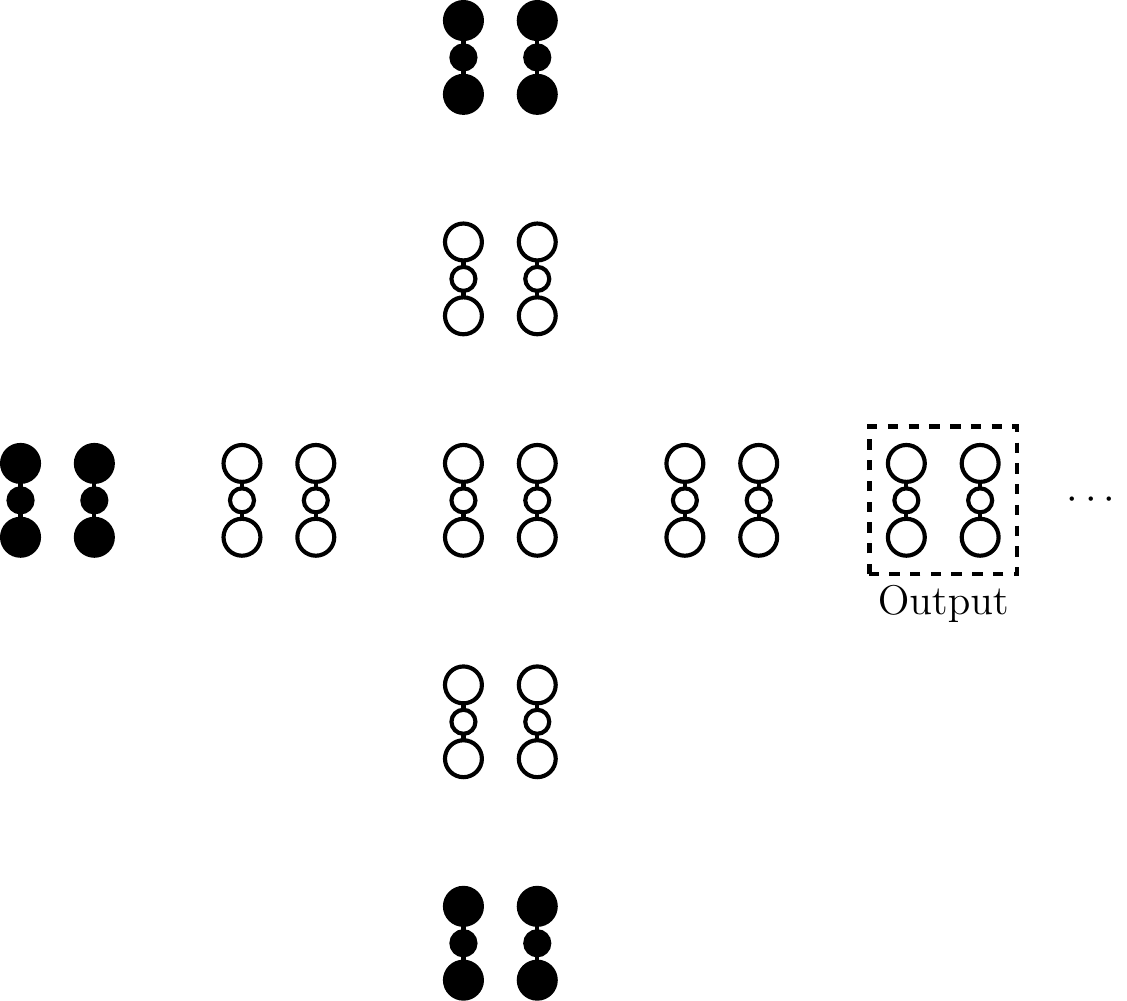}
      }
      \quad
    \subfloat[Input-insensitive majority. \label{fig:majority-imp}]{
      \includegraphics[width=0.45\columnwidth]{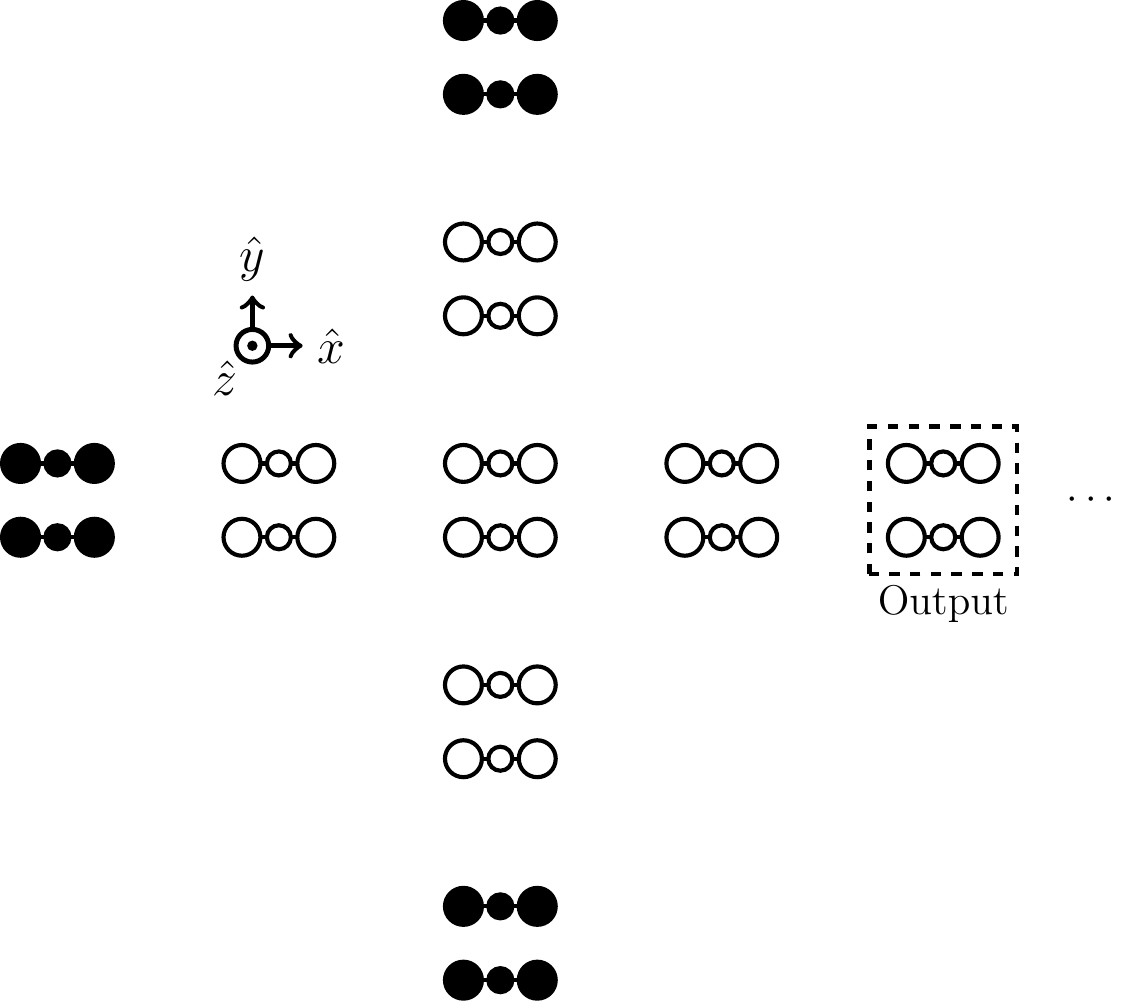}
      }
\caption{(a) A majority gate is built from twelve molecules (six cell pairs). Three driver pairs provide three independent input bits. (b) The cell pairs are rotated 90 degrees to eliminate sensitivity to the input field, \(E_y \hat{y}\). \label{fig:model:majority}}
\end{figure}

Twelve cells (six cell pairs) are used to test the response of a majority gate, as shown in Fig.\ \ref{fig:majority-reg}. 
The three driver cell pairs provide the external driving potential of three independent input bits. Since the coupling strengths between the three input bits and the center device cell group are approximately equal, the device cell pair takes the majority input bit. The majority bit is then shifted downstream by the clock. The second cell pair after the central device cell is taken here to be the output.

\begin{figure}[htbp]
\centering
\includegraphics[width=\linewidth]{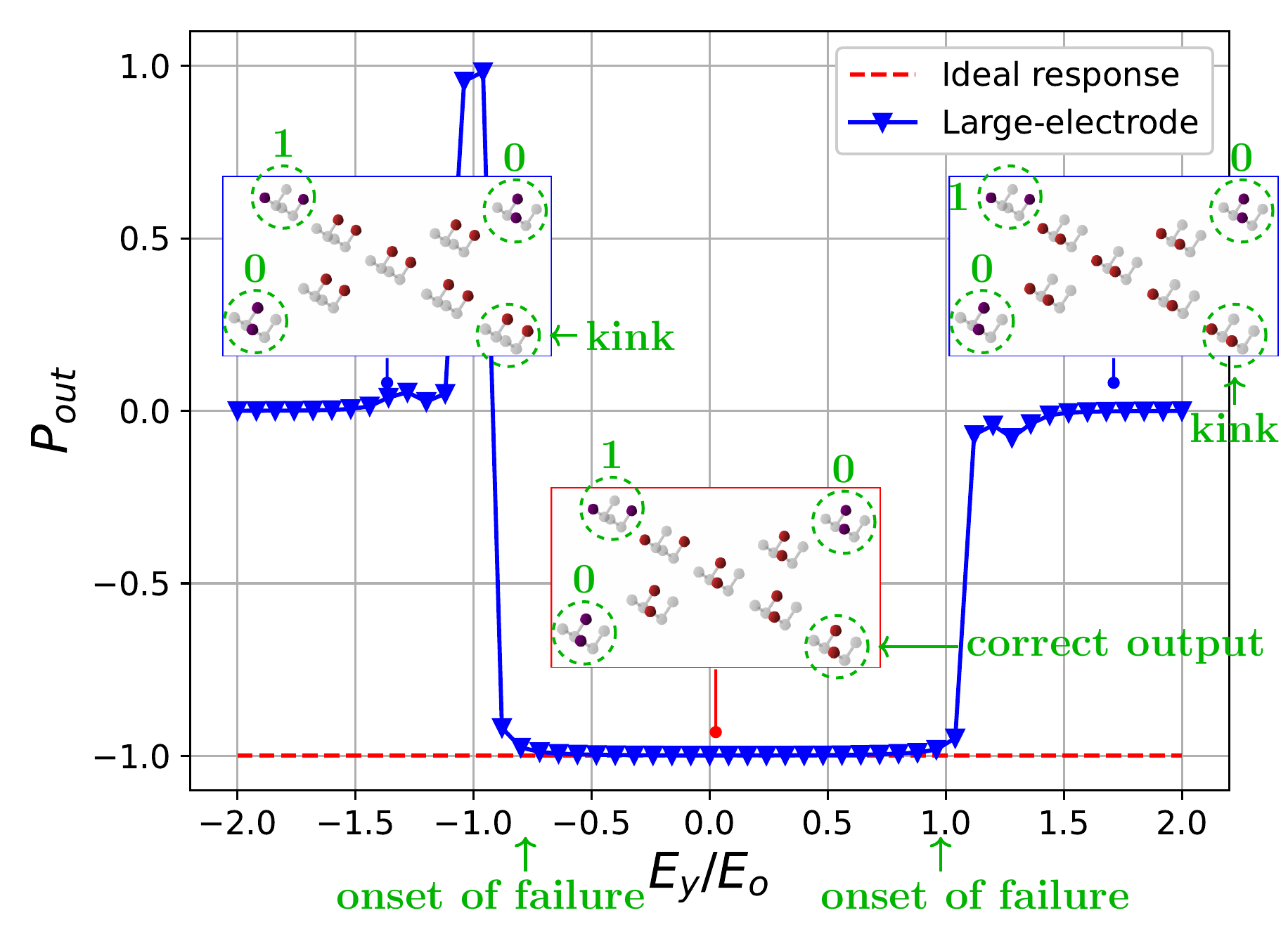}
\caption{A clocked QCA majority gate respond to input bit configuration \(\left(1, 0, 0\right)\) and applied fields in the dominant direction. The majority input bit is \(0\), and the expected output bit is \(0\). \label{fig:majority:result}}
\end{figure}

Fig.\ \ref{fig:majority:result} shows that the ground state of the majority gate correctly encodes the majority of three input bits: \(M(1,0,0) = 0\). The ground state matches the ideal result with \(E_y = 0\) (\(P_{\text{out}} = -1\)) until the field strength in the input direction approaches \(\left| E_y \right| \sim 0.85 E_o\).

\begin{figure}[htbp]
\centering
\includegraphics[width=\linewidth]{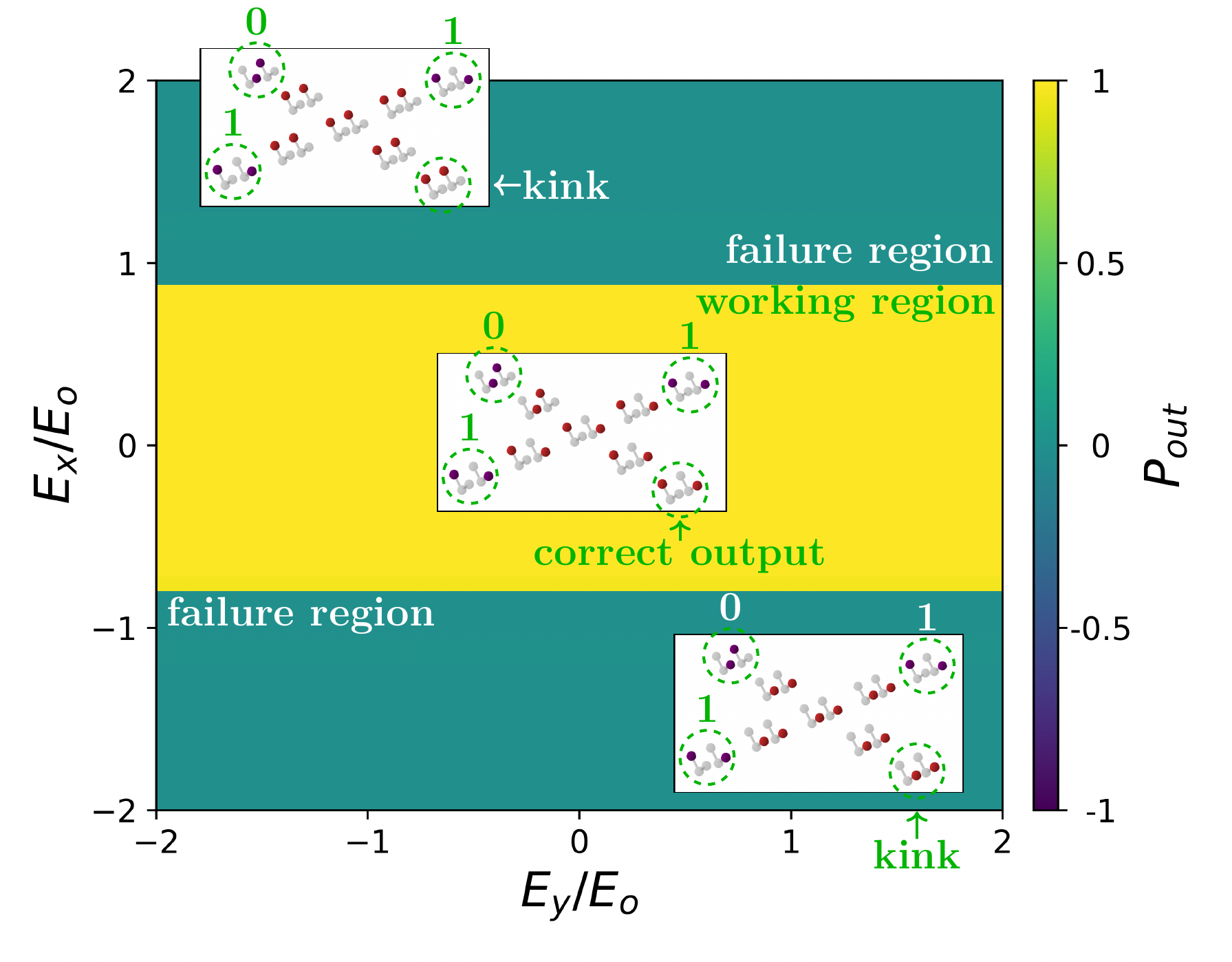}
\caption{A clocked QCA majority gate with rotated cell groups responds to input bit configuration \(\left(0, 1, 1\right)\) and applied non-clocking fields, \(E_x \hat{x} + E_y \hat{y}\). The circuit ground state encodes the correct bit \(M(0,1,1)=1\) regardless of the input field \(E_y \hat{y}\), and for various \(E_x \hat{x}\). \label{fig:majority:result:rot}}
\end{figure}

We demonstrate immunity of the circuit to the input field, \(E_y \hat{y}\), with rotated cell pairs in Fig.\ \ref{fig:majority:result:rot} . Here, the circuit ground state is calculated with input \((0,1,1)\), and the correct output, \(M(0,1,1)=1\) with \(P_{\text{out}} = 1\) may be obtained regardless of \(\left|E_y\right|\). An \(x\) component with strength \(\left| E_x \right| \gtrsim 0.8 E_o\) causes device failure.

\begin{figure}[htbp]
\centering
\includegraphics[width=\linewidth]{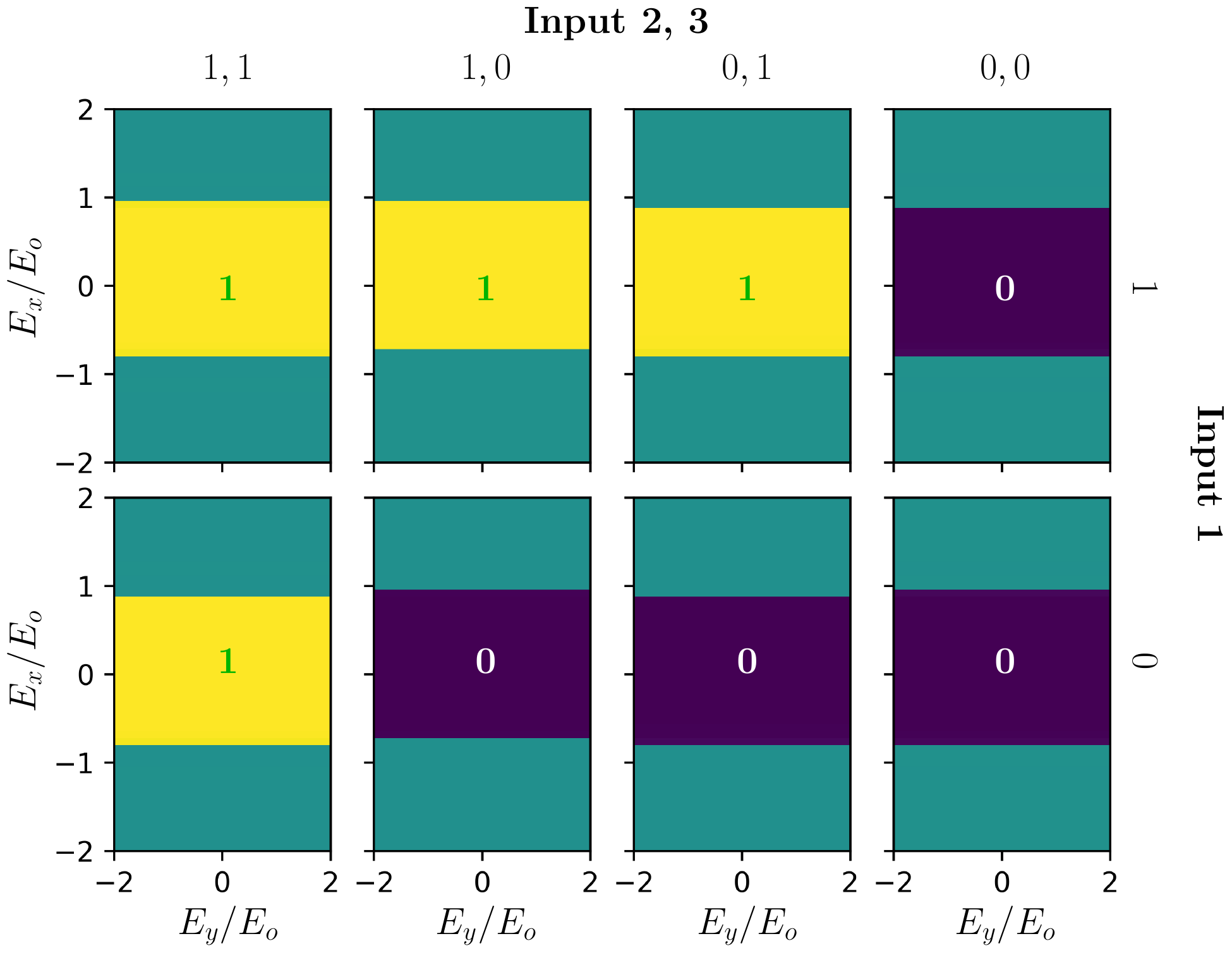}
\caption{A clocked QCA majority gate with rotated cell pairs responds to every possible input configuration and applied non-clocking fields \(E_x \hat{x} + E_y \hat{y}\). Each of the eight subplots  represents a unique input combination. These are arranged in a Karnaugh map. The majority is insensitive to arbitrarily strong \(E_y \hat{y}\) and tolerates strong \(E_x \hat{x}\). \label{fig:majority:result:kar}}
\end{figure}

Like the other circuits, the  majority gate with rotated cell pairs is unaffected by \(E_y \hat{y}\) and tolerates strong \(E_x \hat{x}\).
The circuit responses to all \(2^3\) input bit configurations under applied electric fields \(E_x \hat{x} + E_y \hat{y}\) is shown in Fig.\ \ref{fig:majority:result:kar}. The various input combinations are arranged in a Karnaugh map. In each case, the circuit returns the correct output bit, despite a strong unwanted \(E_x \hat{x}\) up to \(\left|E_x\right| \sim 0.8E_o\). As discussed earlier, disruptive field strengths \(\left|E_x\right|\) are easy to avoid. Thus, the majority gate and the basic logic devices modeled here may be very robust and tolerate significant unwanted non-clocking field components, \(E_x \hat{x} + E_y \hat{y}\).

\section{Conclusion \label{conclusion}}
\label{sec:org0c5ede5}

We have seen that the elementary QCA logic gates and interconnective circuits tolerate significant unwanted, applied non-clocking electric fields, \(E_x \hat{x} + E_y \hat{y}\). While such fields could be from numerous sources, one of the most significant may be input electrodes used to write bits to input circuits. Here we have studied the circuits in a worst-case, large-electrode limit in which the full strength of an input field \(E_y \hat{y}\) is applied over QCA logic. It is seen that the logic and interconnects tolerate very strong input fields, \(\left|  E_y \right| \lesssim 0.8 E_o\), while still encoding the correct logical output in the circuit ground state. These circuits are indeed very robust, since even very weak inputs \( \left| E_y \right| \ll 0.5 E_o\) are sufficient to write bits onto input circuits. This means that we may write bits to input circuits using relatively weak input fields strengths \(\left| E_y \right|\), avoiding by a significant operating margin those field strengths that approach disruptive levels.

Additional measures may be taken to minimizing the likelihood of disrupting logical operations with fringing fields from input electrodes. First, if \(E_y \hat{y}\) is used as the input field component, for example, then cells used for interconnects and logic may be rotated such that \(\vec{a} \cdot \hat{y} = 0\), making the cells insensitive to the field component which is likely to dominate. While this introduces sensitivity to the other non-clocking field component, \(E_x \hat{x}\), the circuits demonstrate the same level of tolerance to this component, which will likely be weaker than \(E_y \hat{y}\). Next, spatial separation between input and logic may be used to suppress the fringing input fields over the logic gates. Finally, a phase delay may be used in the clock so that input and logic are operated out of phase: the logic may be either latched or clocked to the ``Null'' state while input fields \(E_y \hat{y}\) are applied; and input electrodes may be de-energized while logical operations are being calculated and logic is being activated from ``Null'' to active states.

\bibliography{QCAReferences}

\end{document}